S.V. Sintsov,* A.V. Vodopyanov, M.E. Viktorov, M. V. Morozkin and M. Yu. Glyavin.

*Non-equilibrium atmospheric-pressure plasma torch sustained in a quasi-optical beam of subterahertz radiation.*

Institute of Applied Physics RAS, 46 Ul'yanov Street, 603950, Nizhny Novgorod, Russia

sins@ipfran.ru *

ORCID 0000-0003-2146-3189

*Abstract*

This paper studies a continuous atmospheric-pressure discharge maintained by focused subterahertz radiation of a 263-GHz gyrotron. A plasma torch propagating towards microwave radiation was sustained on an argon flow exiting a metal gas tube in the surrounding atmosphere of air. Using a high-speed camera, the spatial structure of the torch was studied and the temporal dynamics on timescales of the order of 20 ns was examined. The emission spectra of the optical wavelength range were used to estimate the excitation temperature of argon atoms and electron density. It was shown that a discharge of this type is substantially non-equilibrium and the electron temperature is many times higher than the vibrational and gas temperatures, being 1.5–1.7 eV in order of magnitude. The electron density measured by the Stark broadening of the Hα and Hβ hydrogen lines exceeds a critical value for the frequency of the heating field and is at a level of $1.5 \cdot 10^{15}$ cm$^{-3}$.

*Introduction*

In recent years, sources of high-power radiation of the subterahertz range [1-5] have been rapidly developed. The creation of such high-power sources opens up wide opportunities for studying the properties of this previously inaccessible frequency range. First of all, it is of interest to use such sources of microwave radiation to sustain plasma. Due to the high radiation frequency, it is possible to create gas discharges with high densities in the range $10^{14}$–$10^{17}$ cm$^{-3}$ [6-9]. Due to quasi-optical focusing systems, it is possible to create Gaussian beams of subterahertz radiation with high power densities at a level of 1–100 MW/cm$^2$ and therefore organize a high specific energy input into a gas discharge [10-12]. In the subterahertz and terahertz ranges, the energy input into the discharge is performed by heating the electron component, which makes it possible to create a plasma with unique non-equilibrium characteristics that are not accessible when the plasma is heated by other methods.

At present, low-pressure gas discharges maintained in quasi-optical Gaussian beams of subterahertz radiation are being intensely studied. For example, in the papers by [8, 9, 12] a gyrotron system with a radiation frequency of 670 GHz and a pulse power of up to 40 kW was used as the radiation source. Due to the high heating frequency in such discharges, it is possible to obtain a plasma with density critical for the frequency of the heating field and a high degree of ionization. The achieved plasma parameters contribute to the efficient creation of multiply charged ions, which are effective sources of extreme ultraviolet radiation [13, 14]. The energy input into UV radiation can reach 10% of the input power, which makes this type of discharge attractive as sources of extreme UV for high-resolution lithography [15, 16].

A number of papers based on the Novosibirsk free electron laser (NovoFEL) as a unique source of terahertz radiation with a frequency of 2.3 THz [17, 18] are known. The atmospheric-pressure discharge maintained in a quasi-optical beam of terahertz radiation is one of the research areas [19]. A plasma density exceeding two times a critical value for the frequency of the heating field ($1.5 \cdot 10^{17}$ cm$^{-3}$) and an electron temperature at a level of 2.5 eV were achieved in such a discharge. The study of such a unique continuous discharge of atmospheric pressure was motivated by its possible use as a source of extreme UV.

Previously, we explored the parameters of a continuous discharge maintained in a quasi-optical microwave beam with a frequency of 24 GHz [20]. The electron temperature measured by two independent methods was obtained at a level of 1–1.2 eV. Due to a significant non-equilibrium in the distribution of temperature characteristics, such a plasma source may be of interest in a wide range of plasma-chemical applications. It was shown in [20] that an increase in the energy input into such a discharge led to a proportional increase in the torch sizes, while the temperature characteristics did not change within the error. The use of a microwave 24-GHz radiation source for heating the atmospheric-pressure plasma imposes certain constraints on the attainable plasma parameters, namely, electron temperature and electron density. The achieved plasma parameters can be improved by increasing the heating frequency.

To this end, we have explored a continuous atmospheric-pressure gas discharge maintained in a quasi-optical beam of subterahertz radiation with a frequency of 263 GHz. Such a discharge of atmospheric pressure is a new object in plasma physics. The study of the parameters of such a plasma torch and its spatiotemporal dynamics is an important fundamental problem in understanding the physics of atmospheric-pressure microwave discharges and interesting for possible future applications.

*1. Description of the experiment*

*1.1. Experimental setup*

Figure 1 shows the experimental setup. The scheme for organizing a gas discharge is similar to that described in [6, 20]. A 263-GHz gyrotron with a power of up to 1 kW in continuous mode was used as a source of microwave radiation [21]. The generated microwave radiation has a Gaussian intensity distribution in the beam cross section at the output. Through a quasi-optical path, the radiation was introduced into the gas discharge chamber, where it was focused using a parabolic mirror. The total angular spread of the beam is 60 deg and the beam waist is 1.2 mm. This corresponds to a power density of up to 20 kW/cm$^2$ in the waist and a r.m.s. electric field of 2.7 kV/cm [6, 20]. A metal tube with an inner diameter of 3 mm, along which a plasma-supporting gas was fed towards the propagating field, was placed in the beam waist region. The beam area in the waist region coincided with the cross section of the gas flow at the output of the tube.

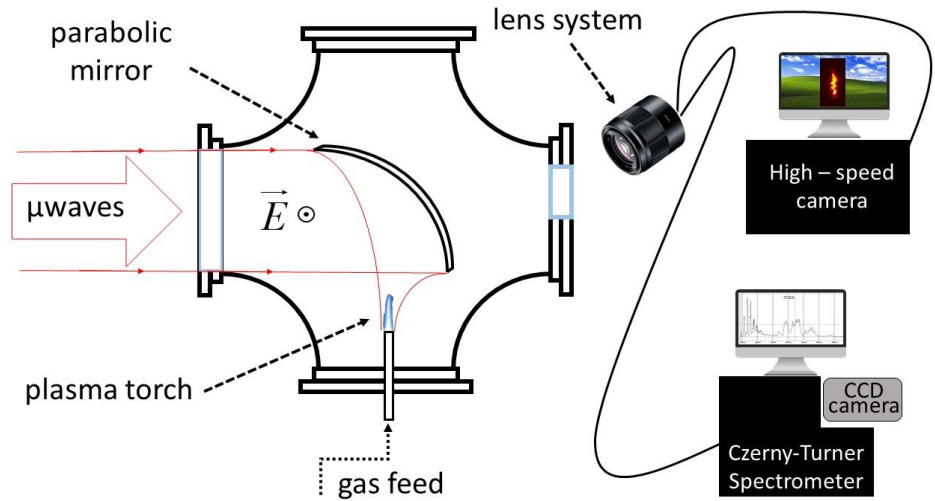

Figure 1. Layout of the experimental setup.

In this paper, argon was utilized as the plasma-supporting gas. The discharge was triggered in an argon stream flowing out towards the propagating field. The discharge is a plasma torch with a diameter equal to the diameter of the gas tube and a length of 1.5–2 cm (see Fig.1). The argon torch burned in the surrounding atmosphere of air at atmospheric pressure. The argon flow rate could be varied from 5 to 30 l/min, which corresponds to a gas flow rate of 12–72 m/s at the pipe section. The discharge was maintained at various input powers of 520 to 1000 W. Visually, the size of the torch almost did not change when the maintenance mode was changed.

Diagnostic flange inputs were provided in the gas-discharge chamber for shooting the discharge on a high-speed camera and recording its optical spectra.

*1.2. High-speed photo of a plasma torch*

To study the spatial structure of the discharge and its temporal dynamics, photographs of the discharge were obtained. For this, we used a high-speed camera with a NanoGate 24 electronic shutter, which made it possible to obtain frames with a minimum exposure of 20 ns. Photos of the discharge were obtained for various modes of sustaining the torch, namely, for different powers of microwave radiation that creates the plasma and argon flows. Figure 2 shows typical discharge photographs taken with exposures of 20 and 1000 ns. In all photographs, the gain was selected in such a way as to ensure the maximum dynamic range of the detected optical plasma glow. The color scale is presented on a logarithmic scale in order to best demonstrate the contour and structure of the discharge for different experimental parameters.

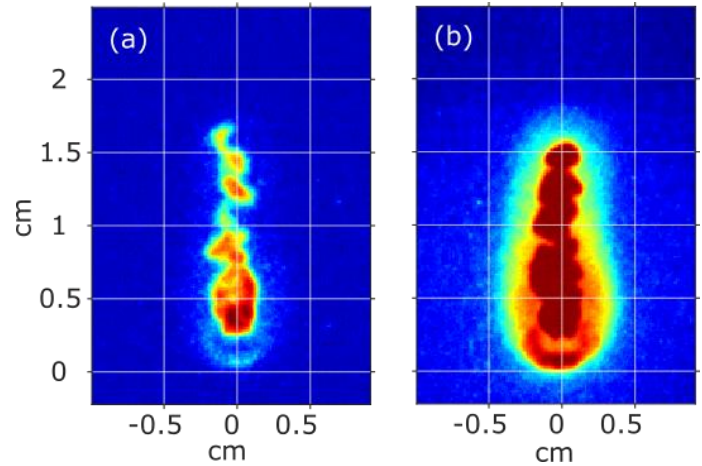

Figure 2. Photos of the discharge on a high-speed camera. The heating power is 1 kW and the argon flow rate is 10 l/min: (a) exposure time 20 ns, (b) exposure time 1 μs.

On a photograph with an exposure time of the order of 1 μs the torch represents a homogeneous plasma structure. The shape and size of the torch do not change on these exposure timescales. On timescales of about 20 ns, the plasma torch is an inhomogeneous structure with quasi-periodic luminosity islands. The characteristic scale of such islands and the distance between them are 1–2 mm. Figure 3 shows several photographs of the discharge with an exposure time of 20 ns taken for one plasma maintenance mode. It can be seen that such an inhomogeneous torch structure is unsteady with time. Luminosity islands change their position in space from photograph to photograph. It can be seen in Fig. 3 that the lower part of the discharge adjacent to the gas tube is fixed. This region has the form of a ball with a characteristic diameter of the order of the diameter of the gas tube. The shape and position of this luminous region do not change. We will call this region the discharge body.

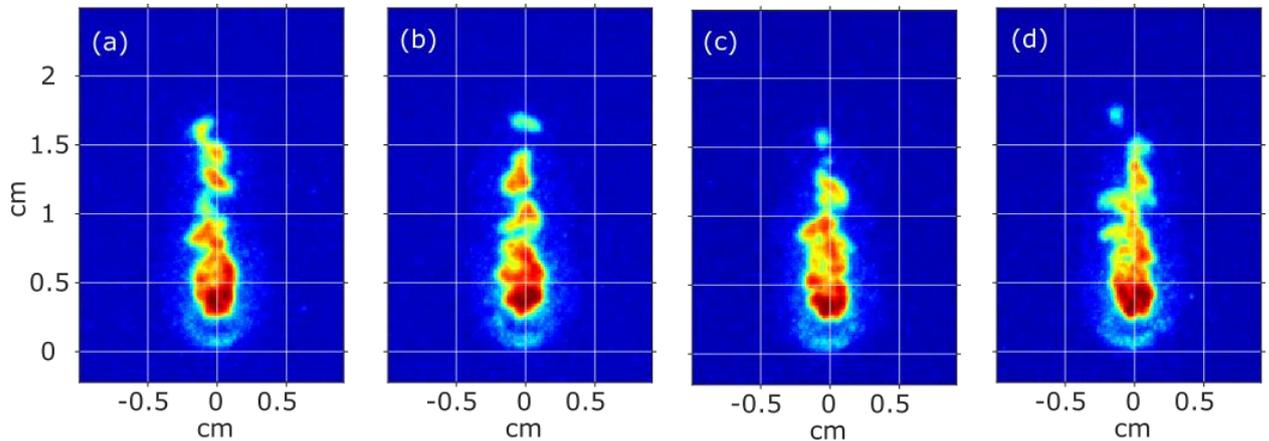

Figure 3. Photos of the discharge on a high-speed camera. The heating power is 1 kW, the argon flow rate is 10 l/min, and the exposure time is 20 ns. The discharge body as the unchanging part is marked on the photographs.

Figure 4 shows discharge photographs taken for various values of the heating power with other conditions being equal. It can be seen that the total size of the torch decreased with decreasing heating power. The size of the torch body also decreased. Based on the analysis of numerous high-speed photographs taken for various modes of maintaining the discharge, it was found that the size of the torch body varies in proportion to the power of the heating field.

The discharge region rising above the discharge body is drifted by an argon flow towards the incident quasi-optical microwave beam. The position of the luminosity boundary of this plasma rising from the discharge body is determined by the waist power density sufficient to maintain the discharge. The length at which the beam power density is such that it is not able to keep a sufficient ionization rate will be exactly the torch length that we see with exposure times of more than 1 μs.

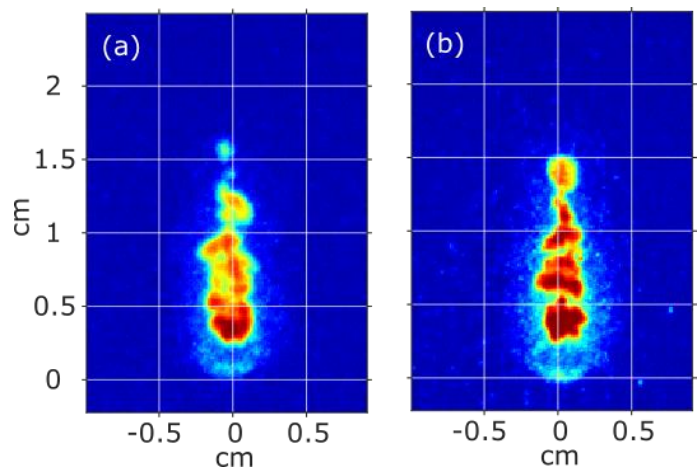

Figure 4. Photos of the discharge on a high-speed camera: (a) heating power 1 kW, argon flow rate 10 l/min, and exposure time 20 ns, (b) heating power 0.78 kW, argon flow rate 10 l/min, and exposure time 20 ns. The region marked on the photographs is the discharge body.

The transfer time of such island structures in the plasma torch volume is much less than 1 μs, since during this time there is complete averaging of the luminosity pattern. The argon flow rate in the experiments ranged from 12 to 72 m/s. This means that such a transfer cannot be caused by gas-dynamic processes. Otherwise, such a stationary form of the torch would have to be observed only on timescales of the order of 1 ms. Therefore, such an inhomogeneous quasi-periodic structure of the plasma glow can be due to the structure of the microwave field in the plasma maintenance region. Indeed, due to the reflection of the incident microwave radiation from the torch body, an unsteady standing field structure [22], the characteristic variation time of which is a few nanoseconds in order of magnitude, arises in the plasma volume [23].

*1.3. Measurement of optical emission spectra*

To study the parameters of a plasma torch sustained by the radiation of a 263-GHz gyrotron, optical emission spectra were obtained. Optical spectra were recorded using a Czerny–Turner monochromator MS3205i (manufactured by SOL Instrument). The resolution of this monochromator for a wavelength of 546 nm with a grating of 1200 lines per millimeter and a slit width of 10 μm was 0.028 nm. A highly sensitive iCCD matrix was used as a detector in this spectrometer. For this configuration, the width of the spectrometer hardware function is 0.15 nm, which was determined from the line emission spectrum of a neon lamp. Optical spectra were recorded in the range from 300 to 900 nm. Using a lens system, an image of the torch was obtained, providing an integrated record of the entire plasma volume. For the optical system used in the measurements, the sensitivity of the matrix was calibrated using a SIRS 610 (USSR) calibration lamp. From the blackbody radiation of such a lamp with a temperature of 3000 K, we obtained a normalization table, which was subsequently used to correct the relative intensity of the lines.

Figure 5 shows a typical emission spectrum of a plasma torch. The torch burns in a stream of argon in the surrounding atmosphere of air. Herein, air from the surounding atmosphere of the torch is mixed in a small amount into the plasma volume. Therefore, the emission lines of not only argon, but also nitrogen and oxygen are observed in the obtained spectra. By the shape of a typical emission spectrum shown in the figure, two characteristic regions can conditionally be distinguished, namely, the red region, where intense argon lines are mainly observed, and the blue region, where both argon emission lines and emission electron-vibrational transitions of the second positive nitrogen system are observed. The intensity of argon lines from the red region is almost two orders of magnitude higher than the intensity of nitrogen lines in the blue region.

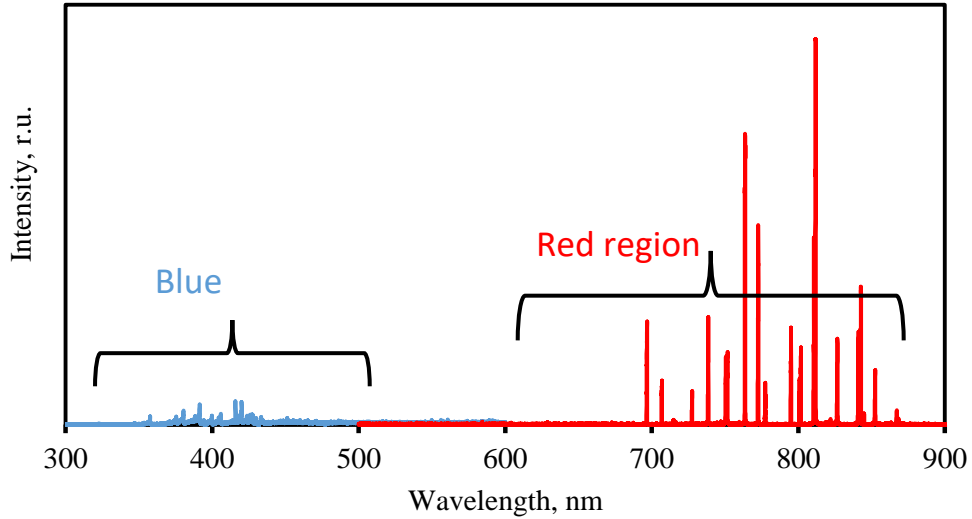

Figure 5. Typical emission spectrum of a plasma torch. The microwave heating power is 1 kW and the argon flow rate is 10 l/min. The exposure time is 6 ms and the slit width is 10 μm.

For the further analysis of plasma parameters, both argon emission lines and nitrogen emission lines were used. Therefore, for each mode of sustaining the torch (with different heating power or gas flow rate), two spectra with different amplification were taken

- in the range 300–900 nm with an exposure time of 6 ms;

- in the range 300–600 nm with an exposure time of 60 ms.

In this paper, based on the obtained optical emission spectra, temperature characteristics of the discharge were estimated within the framework of the partial local thermodynamic equilibrium approximation:

- rotational and vibrational temperatures were estimated from the nitrogen emission lines;

- electron excitation temperature was estimated from the argon emission lines.

## 2. Measurement of temperature characteristics

### 2.1. Measurement of vibrational temperature

The vibrational temperature was estimated by the relative intensity of the emission lines of the second positive nitrogen system. The resolution of the spectrometer was sufficient to isolate individual lines of electron-vibrational transitions of the nitrogen molecule. The estimation was based on the assumption of the Boltzmann distribution of electron-vibrational energy states. To improve the accuracy of the estimates, the vibrational temperature was evaluated independently from the second and third sequences of this line system. The technique for estimating the vibrational temperature from the second positive nitrogen system is widely known and described in detail in [24].

Figure 6 shows the results of measuring the vibrational temperature as a function of the argon flow for different values of the supplied microwave power. The processing included two sequences of electron-vibrational transitions of the second positive nitrogen system. In order of magnitude, the vibrational temperature was 3000 K. The error was determined by the standard deviation of the obtained distribution of vibrational states from the Boltzmann distribution. A common trend can be singled out in the behavior of the vibrational temperature, depending on the power supplied. Vibrational temperature decreases within 10–15% as the argon flow rate is increased.

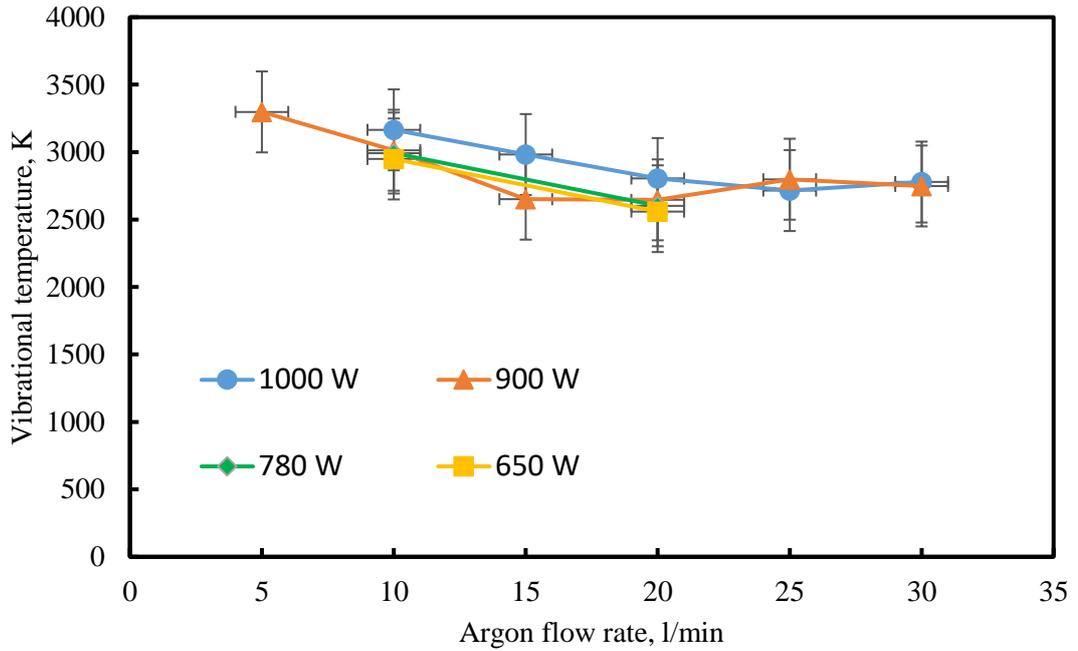

Figure 6. Vibrational temperature of nitrogen as a function of the gas flow rate for different heating powers.

*2.2. Measurement of rotational temperature*

Rotational temperature was also estimated from series of lines of the second positive nitrogen system. The resolution of the spectrometer did not allow resolving the rotational structure of the electron-vibrational transitions of the nitrogen molecule. In this case, the rotational temperature is estimated at the edge of the rotational line system. The estimation technique consists in selecting the value of the rotational temperature corresponding to the best fit of the obtained distribution of vibrational-rotational states to the Boltzmann distribution. The estimation technique we used is described in detail in [25].

Figure 7 shows the obtained dependence of the rotational temperature on the gas flow for different heating powers. In order of magnitude, this quantity is 2000 K and varies within 500 K. The behavior of the obtained dependences is similar to the behavior of the vibrational temperature. Namely, a slight increase in temperature can be observed against the background of the error, depending on the heating power. One can see a decrease in rotational temperature by 1000 K as the argon flow is increased by a factor of 6.

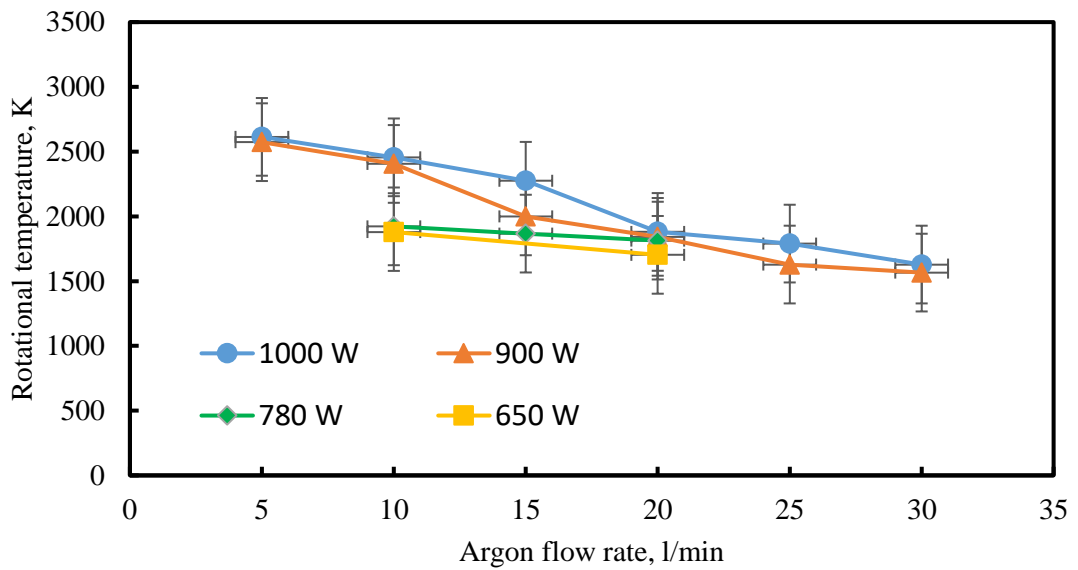

Figure 7. Rotational temperature of nitrogen as a function of the gas flow rate for different heating powers.

Due to the high speed of energy exchange processes between the rotational and translational degrees of freedom of molecules, the rotational temperature is approximately equal to the gas temperature [26]. At least, the obtained values of the rotational temperature are the upper bound for the gas temperature. A number of papers dealing with atmospheric-pressure discharges clearly demonstrate that the degree of difference between rotational and gas temperatures does not exceed 10% [27]. Therefore, in the future, when discussing the non-equilibrium state of this type of discharge, we will speak of rotational temperature as of a gas temperature.

*2.3. Measurement of the argon atom excitation temperature, Electron temperature estimation.*

Using the obtained argon emission lines in the red and blue spectral regions, it is possible to estimate the excitation temperature of argon atoms. This characteristic describes the energy distribution of the excited states of argon atoms. In the literature, such a diagram is called the Boltzmann plot and is used to determine the temperature describing the atomic excitation distribution [29]. Figure 8 gives a view of such a Boltzmann plot obtained in this paper for a discharge maintained by radiation with a power of 1 kW and an argon flow of 10 l/min. For its construction, lines from the red and blue regions of the spectrum were used to increase the reliability of the estimates. The larger the energy gap between the energies of the upper levels, the higher the accuracy of the estimates. More precisely, such an energy gap should be greater than the value of the estimated excitation temperature [30, 31]. Such a technique for estimating the temperature of the excitation of argon atoms is effective provided that the distribution of argon atoms over the excited states is described by the Boltzmann statistics. This statement is not always true, especially in non-equilibrium discharges with features in the electron energy distribution function [32, 33]. In such discharges, the electrons that provide the excitation of neutrals partly reproduce their distribution function on the distribution of neutrals over excitation energies. The level excitation and deactivation mechanisms can also play an important role in the filling of excited levels. The dominance of any of the mechanisms can also lead to a violation of the Boltzmann distribution of atomic excitation.

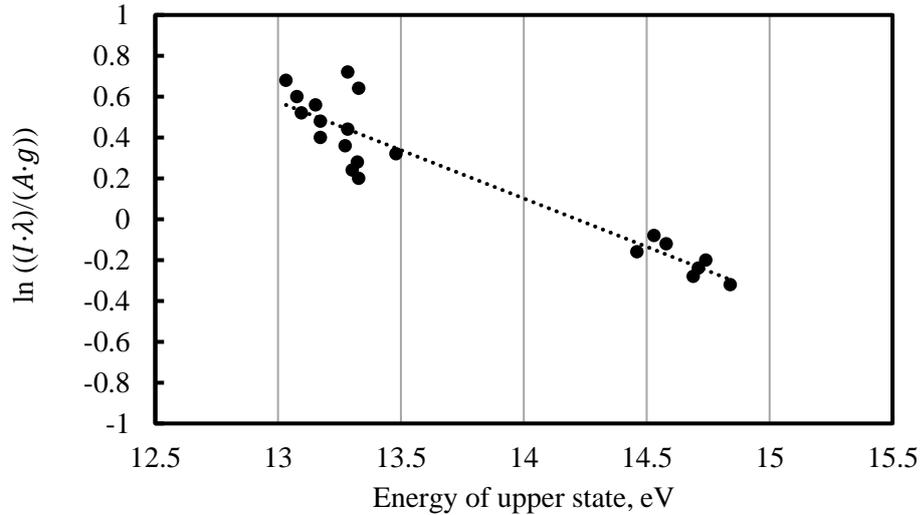

Figure 8. An example of the Boltzmann plot at a mass flow rate of 10 l/min and a microwave power of 1000 W.

In non-equilibrium discharges, where the conditions of partial local thermodynamic equilibrium (pLTE) are often fulfilled, the concept of excitation temperature is introduced not for all levels of excitation energy, but only for some of them, starting with a certain one. The value of this threshold level is determined by the ratio (1) obtained in [23]:

$$n_e \geq 10^{14} \cdot T_e^{1/2} \cdot (E_k - I)^3, \qquad (1)$$

where Ne is the electron density in cm$^{-3}$, Te is the electron temperature in eV, I is the ionization potential in eV, and Eк is the energy of the level at which the distribution starts to become Boltzmann, in eV.

In our case, the estimates based on this formula make it possible to obtain the Boltzmann energy distribution of argon atoms starting with the 11 eV energy level. It can be seen in Fig. 8 that the obtained excitation temperatures conform to the description of the energy distribution of excited argon levels starting with the 13-eV energy level, which indicates their reliability.

The temperature of excitation of a neutral gas in a plasma is directly related to the electron temperature. In the case where the plasma is in the local thermodynamic equilibrium (LTE) state, these temperature characteristics are the same [34]. As the conditions providing the LTE of the plasma become worse, these temperature characteristics gradually diverge. Herein, the electron temperature exceeds the excitation temperature of the energy states of the neutrals. The relation between the electron temperature and the excitation temperature for a plasma in the pLTE state is discussed in [35]. It was shown that the electron temperature can significantly exceed the excitation temperature measured according to the Boltzmann plot. The degree of discrepancy between the electron temperature and the excitation temperature is governed by the plasma parameters and the dynamics of the filling of the upper energy levels.

In the studied type of discharge, the excitation temperature of argon atoms was specified in different modes of plasma maintenance. The obtained dependences are shown in Fig. 9. In order of magnitude, the excitation temperature was 1.5–1.7 eV. The error was determined by the standard deviation of the straight line in the Boltzmann plot (see Fig. 8). It can be seen that the excitation temperature does not depend on the argon flow. One can see a slight increase in the excitation temperature as the heating power is increased.

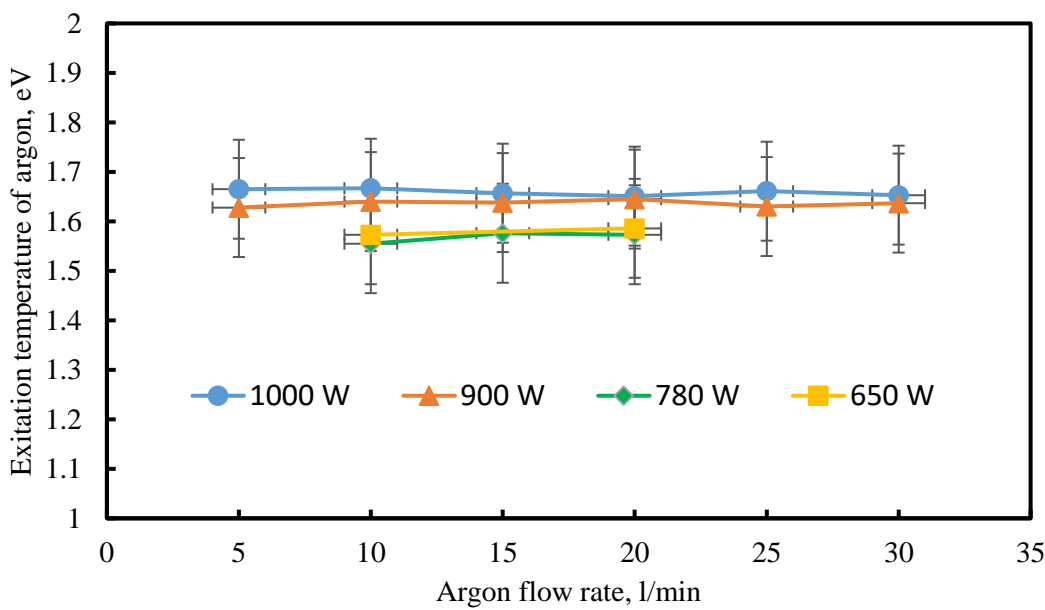

Figure 9. Excitation temperature as a function of the argon flow for different heating powers.

Thus, the argon atom excitation temperature obtained within the framework of this paper is a lower bound for the average electron energy in the type of discharge under study. It amounted to 1.5–1.7 eV. These values will be further used for estimates as the electron temperature.

*2.4. Discussion of temperature characteristics*

Based on the analysis of high-speed photos of the discharge, it can be seen that an increase in the heating power leads to an increase in the geometrical dimensions of the torch body. According to estimates, the volume of the torch body burning at the exit of a gas tube is directly proportional to the power of the heating field. Therefore, the vibrational and rotational temperature characteristics change only slightly as the power introduced into the plasma is increased. The excitation temperature of argon atoms, and hence the electron temperature, slightly increase with increasing absorbed power. The excitation temperature is 1.5–1.7 eV. This level characterizing the average value of the electron energy can be due to the presence of inelastic collisions of electrons with the molecules of the atmospheric gases [23].

Previously, a study was made of the parameters of a plasma torch sustained in a beam of focused radiation of a 24-GHz gyrotron [20]. For that discharge maintained with an order of magnitude lower frequency, the fact of non-equilibrium temperature characteristics was also demonstrated. It was found that the electron temperature in that type of discharge is 1–1.2 eV and does not change as the maintenance parameters are varied. The same dynamics of electron temperature was also obtained for a plasma torch sustained by microwave radiation with a frequency 10 times greater.

This suggests that for such atmospheric-pressure discharges maintained in quasi-optical microwave beams on an argon flow in the surrounding atmosphere of air, there is an energy threshold for the electron temperature with available power densities. This threshold is determined by inelastic collisions of electrons with the molecules of atmospheric gases. For example, for a nitrogen molecule, such an energy threshold is equal to the energy threshold of excitation of vibrations in the ground state, the cross section of which starts with 0.3 eV and is the maximum at the 2–3 eV level. A change in the heating power in this case leads to a change in the specific energy input into the discharge, which in balanced energy ratios leads to a small shift in the average electron energy relative to the energy threshold of inelastic collisions. Such a temperature dependence was obtained in this paper as well (see Fig. 9). A change in the argon flow can undoubtedly affect the degree of mixing molecular gases into the plasma torch volume. However, estimates of the energy balance show [23] that the process of inelastic collisions of electrons, for example, with nitrogen, remains dominant in such plasma parameters over a wide range of nitrogen concentrations (from 10–3% to 10%). Thus, a change in the argon flow did not lead to a change in excitation temperatures, and hence did not affect the electron temperature.

As the argon flow is increased from 5 to 30 l/min, a decrease in both rotational and vibrational temperatures is observed. The vibrational temperature decrease does not exceed 15%, and the rotational temperature reduces by 40%. For fixed values of the heating power and geometrical dimensions of the torch (it was previously shown that these characteristics are interconnected), that is, for a fixed specific bulk energy input, an increase in argon flow leads to a decrease in temperature. This fact is obvious and follows from the thermodynamic balance.

It was previously discussed that the obtained values of the rotational temperature are an upper bound for the gas temperature in the torch. Specific temperature for the excitation of atomic states of argon is a lower bound for the electron temperature. It can be seen that the excitation temperature of argon atoms is a factor of 8 to 11 higher than the estimated gas temperature. This confirms the fact that this type of atmospheric-pressure discharges maintained by high-power subterahertz radiation is not in equilibrium.

*3. Measurement of electron density from Stark broadening of hydrogen lines*

*3.1. Applicability conditions*

The technique for estimating the electron density in a plasma by the broadening of emission lines relative to their natural edge is based on the Stark effect in the Holzmark ion field. For hydrogen-like atoms, the magnitude of the broadening of emission lines is directly related to the density of plasma ions [36]. In the case of a singly ionized plasma, the ion density is approximately equal to the electron density. That is, knowing the magnitude of the line broadening due to the Stark effect, one can obtain an estimate for the electron density in a plasma. This estimate gives reliable results at a plasma density of $10^{14}$ cm$^{-3}$ [37].

A significant constraint on the use of this technique can be introduced by the width of the spectrometer hardware function. In our case, the width of the spectrometer hardware function is less than the width of the measured lines, which permits one to estimate the electron density. Despite this, the procedure of extracting the true contour of the line from that measured against the background of the hardware function is a separate problem. This will be discussed in what follows.

Other mechanisms can also contribute to the broadening of the emission lines relative to the natural edge. These contributions can greatly vary depending on the discharge environment. In particular, the Doppler and Van der Waals broadening mechanisms can make the largest contribution. Based on the above estimates of the gas temperature, it was shown that for this type of discharge, these broadening mechanisms contribute no more than 5%. These effects will not be taken into account in the further analysis. Details about these estimates described in [38].

*3.2. Results of the electron density measurements*

To measure the electron density by the Stark broadening of hydrogen lines, 3 volume percent of hydrogen were added to the plasma-supporting gas (argon). Such an additive did not affect the discharge burning. However, the Balmer-series hydrogen lines appeared in the discharge emission spectrum. The intensity of these lines compared with the noise level was sufficient for analysis of the line broadening. In this paper, by analyzing the radiation of the whole plasma torch, we found the average values of the parameters of the whole plasma volume. Therefore, it was necessary to make sure that the added hydrogen in the plasma-supporting gas is uniformly distributed and excited over the entire plasma volume. For this, a special optical filter with a passband of 656.3±3.5 nm was employed. This range includes the radiation of the Hα hydrogen line of the Balmer series with a wavelength of 656.3 nm. There are no other emission lines in the region of the passband of such an optical filter. Figure 10 shows two photos of the discharge. The left-hand photo was taken without the filter. The right-hand photo was taken through such an Hα filter. In other words, the right-hand photo shows the distribution of excited hydrogen atoms in the torch volume.

It can be seen in Fig. 10 that the luminosity of the excited hydrogen atoms is uniformly distributed over the torch volume. This means that an analysis of the hydrogen lines with a further calculation of the electron density characterizes the plasma parameters of the entire discharge.

For a better resolution of hydrogen emission lines, the exposure time was set to 500 ms. Spectra were recorded for the same discharge burning modes as in the electron temperature measurement.

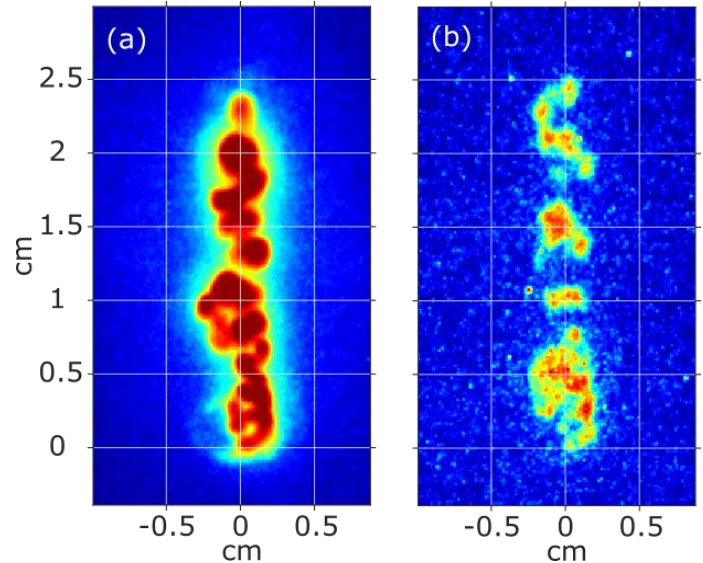

Figure 10. High-speed photo of a plasma torch: (a) without the filter (the exposure time is 1000 ns), (b) through an optical Hα filter with the bandwidth 7 nm (the exposure time is 10000 ns). The power is 1000 W, the argon flow rate is 10 l/min.

According to the emission spectra of the torch, it was found that the characteristic half-width of the Hα line lies in the range 0.18-0.2 nm and the width of the Hβ line is 0.46-0.49 nm. The contour of the measured lines is a convolution of the true contour of the line, which has a Lorentz shape, and of some Gaussian profile, which is due to the spectrometer hardware function [39]. Therefore, the measured profile is described by the Voigt function - a convolution of the Lorentz and Gaussian statistics. To retrieve the true counter line, it is necessary to deconvolute the measured signal with the spectrometer hardware function [40]. A number of approximation formulas are described in the literature to recalculate the half-width of the true line profile from the half-width of the measured line and the half-width of the spectrometer hardware function. Such formulas permit one to significantly simplify the deconvolution of the contours and accurately recalculate the half-width of the true signal. Using the approximation formulas from [36] for the Hα and Hβ hydrogen lines, we calculated the true values of the line half-widths in this paper. The half width of the Hα line lies in the range of 0.05–0.08 nm and the width of the Hβ line is 0.39–0.42 nm.

Based on the recalculated true values of the half-width of the Balmer-series hydrogen lines, we estimated the electron density in the plasma torch. For this, formulas from [37] were also used. It was shown in [38] that the calculations based on such formulas give an estimation accuracy no worse than 10% for Hα and no worse than 5% for Hβ in the existing range of electron densities. Figure 11 shows the obtained values of the electron density in the plasma torch versus power for different flows of a plasma-supporting mixture. Dependences for estimation by the Hα and Hβ lines are separately presented.

It can be seen in the figure that the electron density does not change with increasing power and gas flow within the error. This fact corresponds to the statement made to explain the constancy of the electron temperature. For estimates by the Hα lines, it was found that the average electron density is $1.5 \cdot 10^{15}$ cm$^{-3}$. For estimates by the Hβ lines, the average electron density was $2.5 \cdot 10^{15}$ cm$^{-3}$. Such a difference in the electron density estimates may be due to the inaccuracy in determining the half-width of the true contour of the Hα line, since its measured broadening is close to the half-width of the spectrometer hardware function. Therefore, certain values of the electron density on the Hα line are on the border of applicability of this technique. In this range of electron densities, the measurements by Hβ can be considered more reliable [37]. The characteristic measured values of the half-width of the Hβ line significantly exceed the half-width of the spectrometer hardware function. Thus, it can be said that the electron density in the plasma torch under study is $2 \pm 1 \cdot 10^{15}$ cm$^{-3}$.

Excluding collisions, the cutoff electron density corresponding to heating at a frequency of 263 GHz is 0.86 $10^{15}$ cm$^{-3}$. If we take into account the electron-neutral collision frequency corresponding to the density of a gas with a temperature of 2000 K, then the cutoff electron density will be 0.88 $10^{15}$ cm$^{-3}$. The electron densities measured by the Stark broadening are at a level of 1.5 ± 0.5 $10^{15}$ cm$^{-3}$. Such an exceeding over the electron density of the critical level is characteristic of high-pressure microwave discharges. There are a number of papers where the electron density in microwave discharges is 3–4 orders of magnitude higher than the cutoff density [41,42]. For example, in [43] the electron density in the atmospheric-pressure plasma is at a level of $10^{15}$ cm$^{-3}$ when heated by microwave radiation with a frequency of 2.45 GHz.

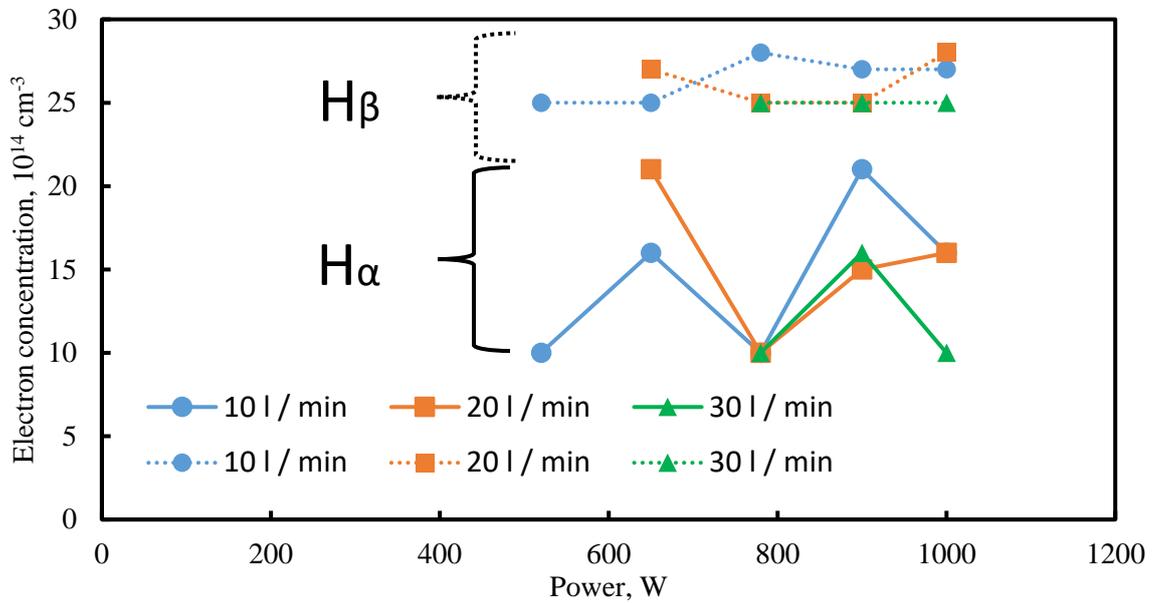

Figure 11. Electron density as a function of power for different flows of a plasma-supporting gas. The results are presented for estimation by the Hα (solid) and Hβ (dotted) lines.

In the studied type of gas discharge, the characteristic size of the plasma torch is 3–4 times greater than the wavelength of the heating field. However, the fact of achieving increased electron density indicates effective penetration of the microwave field into the torch interior. In this regard, it would be interesting to study this type of plasma torch of a smaller size. This can be achieved by decreasing the diameter of the gas nozzle. Reducing the size of the plasma torch can increase the efficiency of plasma heating and provides a different mechanism for input of microwave power [44].

*Conclusions*

We have studied the parameters of the atmospheric-pressure plasma sustained in a quasi-optical microwave beam with a frequency of 263 GHz. The Stark broadening of the Balmer-series hydrogen lines was used to estimate the electron density. The argon excitation temperature was estimated by the argon emission lines. The obtained values are a lower bound for the electron temperature. The emission lines of nitrogen, which is mixed into the plasma torch from the surrounding gas atmosphere, were used to estimate the rotational and vibrational temperatures. The obtained temperature characteristics reliably demonstrate the fact of a non-equilibrium discharge of this type.

Using a high-speed camera with an electronic shutter, we explored the spatial and temporal dynamics of the plasma torch on 20 ns timescales. It was shown that in the region above the gas tube there is a plasma structure, the shape and size of which do not change. The size of such a plasma formation increases in proportion to the input power.

The creation of such non-equilibrium atmospheric-pressure discharges on an argon flow with a high degree of non-equilibrium and a high value of electron densities that exceeds the critical value for the frequency of the heating field allows us to speak of the creation of a novel principle for the organization of plasma-chemical processes. The feed gas, which must be decomposed into certain components, can be fed into the surrounding gas atmosphere of such a non-equilibrium argon torch. The idea is to mix the feed gas into a highly non-equilibrium plasma torch to increase the conversion and efficiency of the plasma-chemical processes.

*Acknowledgments*

This work was supported by the Russian Foundation for Basic Research, projects No. 18-29-21014.